\documentclass[prb,aps,twocolumn,showpacs]{revtex4}

\usepackage{graphicx,epsfig}
\usepackage{bm}

\begin{document}

\title{Strong-to-weak-coupling duality in the
       nonequilibrium interacting resonant-level model}
\smallskip

\author{Avraham Schiller$^{1}$ and Natan Andrei$^{2}$}

\affiliation{$^1$Racah Institute of Physics, The Hebrew
                  University, Jerusalem 91904, Israel\\
             $^2$Center for Materials Theory, Department
                  of Physics, Rutgers University,
                  Piscataway, NJ 08854-8019 USA}

\begin{abstract}
A strong-to-weak-coupling duality is established for
the nonequilibrium interacting resonant-level model,
describing tunneling through a single spinless level,
capacitively coupled to two leads by a contact
interaction. For large capacitive coupling, $U$, and
in the presence of a finite voltage bias, the model
maps onto an equivalent model with a small capacitive
coupling $\rho U' \sim 1/ \rho U$. Here $\rho$ is the
conduction-electron density of states. This duality
is generic to all high-energy cutoff schemes, however
its details may vary from one regularization scheme
to another. In particular, it has the status of an
exact mapping within Abelian bosonization, applicable
to all repulsive interactions whether weak, intermediate
or strong. On a lattice the mapping is restricted to
large $U$, and is controlled by the small parameter
$1/ \rho U$. Explicit expressions are given for the
low-energy scale, differential conductance, and
occupancy of the level in the limit where $U$ is large.
\end{abstract}



\maketitle

\section{Introduction}

The remarkable progress in nanofabrication techniques
has focused considerable experimental and theoretical
interest on quantum impurity systems out of thermal
equilibrium. Typically the impurity is realized by
a small nanostructure, e.g., a semiconductor quantum
dot or an individual molecule, which is attached to
two leads. Application of a voltage bias between the
leads causes steady-state current to flow through the
device. Being strongly correlated, such systems cannot
be studied perturbatively over the full range of
parameters, e.g., temperature or voltage bias, which
calls for the development of nonperturbative techniques
capable of treating the nonequilibrium steady state.

In contrast to the plethora of methods available for
tackling quantum impurities in thermal equilibrium,
only limited progress has been made out of equilibrium.
The difficulty lies in the need to work directly with
open systems, subject to appropriate boundary conditions
imposed by the bias. The latter boundary conditions can
be incorporated either at the level of the many-particle
wave function through solution of a suitable
Lippmann-Schwinger equation (the time-independent
formulation), or by adopting a two-operator description
as in the Keldysh approach~\cite{Keldysh65}
(the time-dependent formulation). Unfortunately,
most of the powerful approaches that are available in
equilibrium are presently inadequate for treating open
systems. In particular, leading numerical methods
such as Wilson's numerical renormalization
group~\cite{Wilson75} or the density-matrix
renormalization group~\cite{White92} are presently
confined to closed finite-size systems.

Recently, a step forward was taken by Mehta and
Andrei,~\cite{MA06} who generalized the Bethe
{\em ansatz} approach to allow the description of
nonequilibrium systems in a steady state. In the
traditional thermodynamic Bethe {\em ansatz} (TBA),
one typically works with closed systems, namely,
finite-size systems subject to periodic boundary
conditions. The limit $L \to \infty$, $L$ being the
size of the system, is taken only at the end of the
construction. This approach, while yielding all the
thermodynamic quantities, does not allow for the
description of the nonequilibrium steady state
that develops in a biased two-lead system, where
particles cross the impurity and dissipate their
energy asymptotically in the leads. A description
of such processes requires different asymptotic
behaviors in the past and in the future, represented
by $x \to \pm \infty$ in the time-independent
formulation. Thus, in the scattering Bethe
{\em ansatz} (SBA) approach~\cite{MA06} one works
directly with open systems, where the limit
$L \to \infty$ is implemented from the outset.
The many-particle scattering states defined on the
open system satisfy the Lippmann-Schwinger equation
with asymptotic boundary conditions of free biased
leads at $x \to -\infty$.

As a first demonstration of the approach, Mehta and
Andrei applied it to the interacting resonant-level
model (IRLM),~\cite{MA06} describing tunneling
through a single spinless level with capacitive
coupling to the leads. The exact SBA solution
revealed several surprising features, including
a nonmonotonic dependence of the current on the
capacitive coupling $U$, and an apparent duality
between $U = 0$ and $U \to \infty$. These results have
stimulated considerable interest in the nonequilibrium
IRLM,~\cite{BVZ07,Doyon07,BS07a,Bohr-Schmitteckert07}
providing complementary points of view and benchmarks
for comparison.

In this paper we provide such a benchmark, by
establishing a strong-to-weak-coupling duality for
the nonequilibrium IRLM. In equilibrium, duality
means that one can map the effective low-energy
Hamiltonians for large and small $U$. Out of
equilibrium the mapping must also preserve the
boundary conditions imposed by the bias, which
sets a far more stringent constraint on the mapping.
In the following we show that duality between strong
and weak coupling is a generic feature of the
nonequilibrium IRLM, independent of the cutoff
scheme used. However, its details may vary from
one regularization scheme to another. This is an
important point to bear in mind whenever comparing
different computational schemes. Indeed, within
Abelian bosonization we find an {\em exact mapping}
between $U$ and $U' = 4/(\pi^2 \rho^2 U)$, where $\rho$
is the conduction-electron density of states per
lattice site. Thus, $U = 0$ and $U \to \infty$ are
equivalent within bosonization, in agreement with the
SBA. A similar duality between $\rho U \gg 1$ and
$\rho U' \sim 1/\rho U \ll 1$ is found on a lattice,
however the mapping is only approximate, and is
accompanied by a significant reduction of the associated
low-energy scale according to $T_0(\rho U \gg 1)
\sim T_0(U=0)/(\rho U)^{2}$. Explicit analytical
expressions are given in the limit $\rho U \gg 1$
for the low-energy scale, $T_0$, the differential
conductance, $G$, and the occupation of the level,
$n_d$.

The outline of the paper is as follows. In
Sec.~\ref{sec:model} we introduce the nonequilibrium
IRLM, and specify our formulation of the finite bias.
Using Abelian bosonization, an exact mapping between
$U$ and $U' = 4/(\pi^2 \rho^2 U)$ is derived in
Sec.~\ref{sec:bosonization}, followed by
treatment of a general lattice cutoff in
Sec.~\ref{sec:lattice-cutoff}. Explicit
analytic expressions are obtained in turn in
Sec.~\ref{sec:expressions} for the low-energy scale,
differential conductance, and occupancy of the level,
in the limit where $U$ is large. Finally, we present
our conclusions in Sec.~\ref{sec:conclusions}.

\section{Nonequilibrium interacting resonant-level model}
\label{sec:model}

As already indicated above, there are several possible
ways to formulate quantum impurity systems out of
equilibrium. Here we adopt a two-operator formalism,
consisting of the Hamiltonian of the unbiased system,
${\cal H}$, and the nonequilibrium operator, $Y_0$,
which specifies the bias. This choice of operators
arises naturally when starting from a coupled system
in thermal equilibrium, and switching on the bias
at some remote time $t_0$. Thus, the system is
described at time $t_0$ by the density operator
$\hat{\rho}_0 = e^{-\beta {\cal H}}/{\rm Tr}\{
e^{-\beta{\cal H}}\}$, and evolves henceforth in
time according to the Hamiltonian of the biased
system, $\bar{\cal H} = {\cal H} + Y_0$. Steady
state is reached if the limit $t_0 \to -\infty$
exists, as can be shown in this case.~\cite{DA06}

An equivalent, more common formulation of steady
state starts from two decoupled leads, each in
thermal equilibrium with its own chemical potential.
Tunneling is then switched on at time $t_0$, which
drives the system to a new steady state. Both
formulations illustrated above give identical
results.~\cite{DA06} However, we shall focus on
the former one, as it affords a more concise
presentation. In the SBA approach, applied directly
in the steady-state limit, the second formulation was
adopted~\cite{MA06} with $\hat{\rho}_{\rm initial}
= e^{-\beta ({\cal H}_0 - Y_0)}/ {\rm Tr} \{
e^{-\beta ({\cal H}_0 - Y_0)} \}$. Here ${\cal H}_0$
describes the biased system with the leads decoupled.

With the former formulation in mind, the nonequilibrium
IRLM is defined in the continuum limit by the Hamiltonian
\begin{eqnarray}
{\cal H} &=& i \hbar v_F \sum_{j = 1, 2}
               \int_{-\infty}^{\infty}
                     \psi^{\dagger}_{j}(x)
                     \partial_x \psi_{j}(x) dx
\nonumber \\
         &+& \epsilon_d d^{\dagger} d
           + \sum_{j = 1, 2} \sqrt{a}\, t_j
                 \left \{
                    \psi^{\dagger}_j(0) d + {\rm H.c.}
                 \right \}
\nonumber \\
         &+& U a \left (
                    d^{\dagger} d - \frac{1}{2}
            \right )
            \sum_{j = 1, 2}\!
                    :\!\psi^{\dagger}_{j}(0) \psi_{j}(0)\!:
\label{H_0}
\end{eqnarray}
and nonequilibrium operator
\begin{equation}
Y_0 = \frac{eV}{2} \int_{-\infty}^{\infty}
      \left [
              \psi^{\dagger}_{1}(x) \psi_{1}(x)
              - \psi^{\dagger}_{2}(x) \psi_{2}(x)
      \right ] dx .
\label{Y_0}
\end{equation}
Here, the left-moving fields $\psi^{\dagger}_{j}(x)$
with $j = 1, 2$ describe free conduction electrons in
lead $j$; $d^{\dagger}$ creates an electron on the
level; $a$ is a short-distance cutoff corresponding
to a lattice spacing; $t_{j}$ is the tunneling matrix
element between the level and lead $j$; $U$ is the
capacitive coupling (contact interaction) between the
level and the leads; and $V$ is the applied voltage
bias. Note that we allow for different tunneling
matrix elements to each of the leads, but the
capacitive coupling is taken to be equal. As for
the fields $\psi^{\dagger}_{j}(x)$, these obey
canonical anticommutation relations
\begin{equation}
\{ \psi_{i}(x), \psi^{\dagger}_{j}(y) \}
              = \delta_{i j} \delta(x - y) ,
\end{equation}
subject to the regularization $\delta(0) = 1/a$.
Hence, the conversion between a Wilson-type
tight-binding representation~\cite{Wilson75} and
the continuum limit is specified by the relation
$c^{\dagger}_{j, 0} = \sqrt{a}\, \psi^{\dagger}_{j}(0)$,
where $c^{\dagger}_{j, 0}$ creates a localized
electron (zeroth Wilson shell) on lead $j$. As for
the symbol $:\!\psi^{\dagger}_{j}\psi_{j}\!:$, it
stands for normal ordering with respect to the
unperturbed Fermi seas of the two leads.

\section{Abelian bosonization}
\label{sec:bosonization}

We first treat the nonequilibrium problem defined
in Eqs.~(\ref{H_0}) and (\ref{Y_0}) using Abelian
bosonization. In the standard fashion,~\cite{Haldane81}
two bosonic fields $\Phi_{j}(x)$ are introduced, one
boson field for each left-moving fermion field. The
fermion fields are written as
\begin{equation}
\psi_{j}(x) = \frac{e^{i\hat{\varphi}}}
                   {\sqrt{2 \pi \alpha}}
              e^{-i \Phi_{j}(x)} ,
\label{fermions-via-bosons}
\end{equation}
where the $\Phi$'s obey
\begin{equation}
\left [ \Phi_{i}(x),
        \Phi_{j}(y) \right] =
        -i \delta_{i j} \pi\, {\rm sign} (x-y) .
\end{equation}
The ultraviolet momentum cutoff $\alpha^{-1} = \pi/a$
is related to the conduction-electron bandwidth $D$
and the density of states per lattice site $\rho$
through $D = \hbar v_F/\alpha$ and $\rho = 1/(2 D)
= \alpha/(2\hbar v_F)$, respectively. The operator
$e^{i\hat{\varphi}}$ in Eq.~(\ref{fermions-via-bosons})
is a phase-factor operator, which comes to ensure
that the different fermions anticommute. Our
explicit choice for $\hat{\varphi}$ reads
\begin{equation}
\hat{\varphi} = \frac{\pi}{2}
          \left [
                   \hat{N}_1 - \hat{N}_2 + 2 d^{\dagger}d
          \right ] ,
\label{phase-operator}
\end{equation}
where $\hat{N}_{j}$ is the total number operator
for electrons in lead $j$. Alternatively, one can
replace the phase-factor operator appearing in
Eq.~(\ref{fermions-via-bosons}) with a Majorana
fermion.

In terms of the boson fields, the Hamiltonian
${\cal H}$ and nonequilibrium operator $Y_0$ take
the forms
\begin{eqnarray}
{\cal H} &=& \frac{\hbar v_F}{4 \pi} \sum_{j = 1, 2}
             \int_{-\infty}^{\infty}
                    (\nabla \Phi_{j}(x))^2 dx
             + \epsilon_d d^{\dagger} d
\nonumber \\
         &+& \sum_{j = 1, 2}
                 \frac{t_{j}}{\sqrt{2}}
                 \left \{
                    e^{i\Phi_{j}(0)} e^{-i\hat{\varphi}}
                    d + {\rm H.c.}
                 \right \}
\nonumber \\
         &+& \delta \frac{a}{\pi^2 \rho}
             \left (
                     d^{\dagger} d - \frac{1}{2}
             \right )
             \sum_{j = 1, 2}\!
                     \nabla \Phi_{j}(0)
\label{H_boson}
\end{eqnarray}
and
\begin{equation}
Y_0 = \frac{eV}{4 \pi}
      \int_{-\infty}^{\infty}
           \left [
                   \nabla \Phi_1(x) - \nabla \Phi_2(x)
           \right ] ,
\label{Y_0_boson}
\end{equation}
where
\begin{equation}
\delta = \arctan
         \left (
                 \frac{\pi \rho U}{2}
         \right )
\end{equation}
is the phase shift associated with $U$. This
bosonized form of the $U$ interaction term stems from
the cutoff scheme used in bosonization.~\cite{ZCJA02}
It follows from the requirement that the exact
scattering phase shift be reproduced in the
limit $t_1, t_2 \to 0$. Importantly, $\delta$ is
bounded in magnitude by $\pi/2$. Although the bosonic
Hamiltonian of Eq.~(\ref{H_boson}) does support
larger values of $|\delta|$, this parameter must
not exceed $\pi/2$ in order for Eq.~(\ref{H_boson})
to possess a fermionic counterpart of the form
specified in Eq.~(\ref{H_0}).

We proceed by converting to new charge and flavor
fields, $\Phi_{c}$ and $\Phi_{f}$, corresponding to
even and odd combinations of $\Phi_1$ and $\Phi_2$:
\begin{eqnarray}
\Phi_{c}(x) &=& \frac{1}{\sqrt{2}}
                \left [
                        \Phi_{1}(x) + \Phi_{2}(x)
                \right ] ,
\\
\Phi_{f}(x) &=& \frac{1}{\sqrt{2}}
                \left [
                        \Phi_{1}(x) - \Phi_{2}(x)
                \right ] .
\end{eqnarray}
In this manner, Eqs.~(\ref{H_boson}) and
(\ref{Y_0_boson}) are rewritten as
\begin{eqnarray}
{\cal H} &=& \frac{\hbar v_F}{4 \pi} \sum_{\nu = c, f}
             \int_{-\infty}^{\infty}
                    (\nabla \Phi_{\nu}(x))^2 dx
             + \epsilon_d d^{\dagger} d
\nonumber \\
         &+& \frac{t_1}{\sqrt{2}}
                 \left \{
                        e^{\frac{i}{\sqrt{2}}
                          [\Phi_{c}(0) + \Phi_{f}(0)]}
                        e^{-i\hat{\varphi}} d + {\rm H.c.}
                 \right \}
\nonumber \\
         &+& \frac{t_2}{\sqrt{2}}
                 \left \{
                        e^{\frac{i}{\sqrt{2}}
                          [\Phi_{c}(0) - \Phi_{f}(0)]}
                        e^{-i\hat{\varphi}} d + {\rm H.c.}
                 \right \}
\nonumber \\
         &+& \delta \sqrt{2} \frac{a}{\pi^2 \rho}
             \left (
                     d^{\dagger} d - \frac{1}{2}
             \right )
             \nabla \Phi_{c}(0)
\label{H_boson-2}
\end{eqnarray}
and
\begin{equation}
Y_0 = \frac{eV}{2 \sqrt{2} \pi}
      \int_{-\infty}^{\infty}
           \nabla \Phi_{f}(x) .
\label{Y_0_boson-2}
\end{equation}
Note that the phase operator $\hat{\varphi}$ of
Eq.~(\ref{phase-operator}) involves only the
flavor field $\Phi_{f}$, and therefore commutes
with $\Phi_{c}$.

At this point two successive transformations are
carried out. First the canonical transformation
${\cal H}' = U {\cal H} U^{\dagger}$,
$Y'_0 = U Y_0 U^{\dagger}$ with $U = \exp \left [
i \sqrt{2} \Phi_{c}(0) \left ( d^{\dagger}d - \frac{1}{2}
\right ) \right ]$ is performed, to be followed by
$\Phi_{c}(x) \to -\Phi_{c}(x)$. The latter operation
is just a bosonic version of the particle-hole
transformation $\psi_1(x) \to \psi^{\dagger}_2(x)$,
$\psi_2(x) \to \psi^{\dagger}_1(x)$. The
nonequilibrium operator $Y'_0$ is left unchanged
by this sequence of steps, and remains given by
Eq.~(\ref{Y_0_boson-2}). The Hamiltonian ${\cal H}'$,
on the other hand, acquires one small but important
modification as compared to Eq.~(\ref{H_boson-2}):
the parameter $\delta$ is replaced with
$\delta' = \frac{\pi}{2} - \delta$. As long
as $|\delta'| \leq \pi/2$, i.e., $U \geq 0$,
one can revert the series of steps leading
to Eqs.~(\ref{H_boson-2}) and (\ref{Y_0_boson-2}), to
recast ${\cal H}'$ and $Y'_0$ in fermionic forms. The
end result of these manipulations is just the original
pair of operators that appear in Eqs.~(\ref{H_0})
and (\ref{Y_0}), with the renormalized interaction
strength
\begin{equation}
U \to U' = \frac{4}{\pi^2 \rho^2 U} .
\label{U'}
\end{equation}
All other model parameters, the voltage bias included,
are unaffected by the mapping. This proves an
{\em exact} strong-to-weak-coupling duality of the
nonequilibrium IRLM in the framework of
bosonization.~\cite{BS07b}

Proceeding to observables, the duality transformation
leaves unchanged the operators describing both the
occupation of the level, $\hat{n}_d = d^{\dagger}d$,
and the current outgoing from lead $j$, $\hat{I}_{j}
= i(e/\hbar) \sqrt{a}\, t_{j} [ d^{\dagger} \psi_{j}(0)
- \psi^{\dagger}_{j}(0) d ]$. Hence the associated
averages, $n_d = \langle \hat{n}_d \rangle$ and
$I = \langle \hat{I}_j \rangle$, obey
the identities
\begin{equation}
n_{d}(T, V; \epsilon_d, t_1, t_2, U) =
n_{d}(T, V; \epsilon_d, t_1, t_2, U') ,
\end{equation}
\begin{equation}
I(T, V; \epsilon_d, t_1, t_2, U) =
I(T, V; \epsilon_d, t_1, t_2, U') .
\end{equation}
In particular, since $U \to \infty$ maps onto $U = 0$,
occupancy of the level and the steady-state current
are identical for $U \to \infty$ to those of the
noninteracting level. This result is in
agreement with that of the SBA,~\cite{MA06} including
the associated low-energy scale being equal to $\Gamma
= \Gamma_1 + \Gamma_2$. Here $\Gamma_j = \pi \rho t_j^2$
is half the tunneling rate to lead $j$.
We emphasize, however, that our mapping extends the
equivalence of $U \to \infty$ and $U = 0$ to all
moments of the current, implying identical full
counting statistics.

The detailed agreement with the SBA suggests that an
analogous duality is present in the latter approach,
in the form of $\pi \rho U_{\rm SBA} \to 1/(\pi \rho
U_{\rm SBA})$. Here the notation $U_{\rm SBA}$ comes
to emphasize that the coupling $U$ is defined within
the SBA cutoff scheme.~\cite{Comment-on-SBA-DOS} The
structure of the duality is reflected in the expression
for the low-energy scale $T_0$, which is given for
$\pi \rho U_{\rm SBA} \leq 1$ by~\cite{MA-unpublished}
\begin{equation}
T_0^{\rm SBA}(\pi \rho U \leq 1) =
          D \left(
                         \frac{\Gamma}{D}
            \right )^{\pi/[\pi + \zeta(\pi \rho U)]}
\end{equation}
with
\begin{equation}
\zeta(x) = 2 \arctan(x) .
\end{equation}
Assuming the SBA duality then yields
\begin{equation}
T_0^{\rm SBA}(\pi \rho U \geq 1) =
          D \left(
                   \frac{\Gamma}{D}
            \right )^{\pi/[\pi + \zeta(1/\pi \rho U)]} ,
\end{equation}
which has the asymptotic form
\begin{equation}
T_0^{\rm SBA}(\pi \rho U \gg 1) \approx
          D \left(
                   \frac{\Gamma}{D}
            \right )^{1 - 2/(\pi^2 \rho U)} .
\label{SBA-T0}
\end{equation}
This form will be examined later in
Sec.~\ref{sec:expressions}.

Thus, the property of duality, Eq.~(\ref{U'}), provides
a simple perspective on the nonmonotonic $U$ dependence
of the low-energy scale $T_0$, discussed in
Ref.~\onlinecite{BVZ07}. Starting from $U = 0$ and
$T_0 = \Gamma$, the low-energy scale increases as a
function of small $U$,~\cite{BVZ07} before dropping
back to its noninteracting value as $U \to \infty$.
It must therefore decrease as a function of large $U$,
attaining a maximum at some intermediate coupling
$U_{\rm max}$. Assuming a single extremum, the maximal
$T_0$ must occur at the self-dual point $U' = U$,
corresponding to $U_{\rm max} = 2/(\pi \rho)$ in
bosonization and $U_{\rm max} = 1/(\pi \rho)$ in the
SBA. Indeed, the estimate $U_{\rm max} \approx 2/(\pi \rho)$
is explicitly found in the Anderson-Yuval
approach.~\cite{Comment-on-AY} Although the
precise value of $U_{\rm max}$ may vary from one
regularization scheme to another, the general
estimate $\rho U_{\rm max} \sim 1$ should apply
to all cutoff schemes, in agreement with the
weak-coupling and numerical renormalization-group
results of Borda {\em et al}.\cite{BVZ07} Finally,
we note that the derivation and structure of
Eq.~(\ref{U'}) resemble a similar duality recently
found in the anisotropic multichannel Kondo
model.~\cite{SdL06}

\section{Lattice cutoff}
\label{sec:lattice-cutoff}

Although the mapping of Eq.~(\ref{U'}) is exact within
bosonization, it does depend on the high-energy cutoff
scheme used. Generally speaking, all regularizations
are expected to share the same qualitative physics.
However details, such as the parametric dependence
of $T_0(U)$, may vary from one realization scheme to
another. It is quite remarkable in that respect that
bosonization and the SBA coincide for $U \to \infty$,
as both approaches employ seemingly unrelated cutoffs.
Focusing on the large-$U$ limit, we proceed to
analyze the strong-to-weak-coupling duality for a
general lattice cutoff. Here by lattice cutoff we
mean an arbitrary conduction-electron dispersion
$\epsilon_k$ with a large but finite bandwidth, $D$.

Consider a general lattice model for the conduction
electrons in the leads. Using a Wilson-type
construction,~\cite{Wilson75} any lattice model
can be cast as a semi-infinite tight-binding
chain with the impurity coupled to the open end:
\begin{eqnarray}
{\cal H} &=&
       \sum_{j = 1, 2}
       \sum_{n = 0}^{\infty}
             \left [
                     \epsilon_n c_{j, n}^{\dagger}
                                c_{j, n}^{}
                     + \xi_n \left \{
                                c^{\dagger}_{j, (n + 1)}
                                c_{j, n}^{} + {\rm H.c.}
                             \right \}
             \right ]
\nonumber \\
   &+& \epsilon_d d^{\dagger}d +
       \sum_{j = 1, 2}
             t_j
             \left \{
                      c^{\dagger}_{j, 0} d + {\rm H.c}
             \right \}
\nonumber \\
   &+& U \left (
                 d^{\dagger} d - \frac{1}{2}
         \right )
         \sum_{j = 1, 2}
         \left (
                  c^{\dagger}_{j, 0} c^{}_{j, 0} -
                  \frac{1}{2}
         \right ) .
\label{H_latt}
\end{eqnarray}
Here $c^{\dagger}_{j, n}$ creates an electron in the
$n$th Wilson shell of lead $j$. The nonequilibrium
operator is given accordingly by
\begin{equation}
Y_0 = \frac{eV}{2}
      \sum_{n = 0}^{\infty}
            \left [
                     c_{1, n}^{\dagger} c_{1, n}^{}
                     - c_{2, n}^{\dagger} c_{2, n}^{}
             \right ] .
\label{Y_0_latt}
\end{equation}
Different lattice models are distinguished by the
tight-binding parameters $\epsilon_n$ and $\xi_n$,
which are uniquely determined by the underlying band
structure and the microscopic details of the coupling to
the level. For instance, the case where $\epsilon_n$ is
zero for all $n$ corresponds to particle-hole symmetric
bands. The bandwidth $D$ is determined in this case
by the largest hopping matrix element $\xi_n$ along
the chain. For convenience, we restrict attention
hereafter to $\epsilon_n = 0$, taking the leads to
have identical forms. As commented on below, both
assumptions can be relaxed without altering the
main physical result.

Focusing on $U \gg D, t_j, |\epsilon_d|$, we carry out
a systematic expansion in $1/U$. For large $U$, it is
favorable energetically to first satisfy the large
capacitive coupling between the level $d^{\dagger}$
and the two local lead electrons $c^{\dagger}_{j, 0}$
with $j = 1, 2$. Diagonalization of the local problem
with $t_1$ and $t_2$ set to zero reveals two low-lying
states:
$c^{\dagger}_{1, 0} c^{\dagger}_{2, 0} |0 \rangle$
with energy $-U/2$, and $d^{\dagger} |0 \rangle$ with
energy $-U/2 + \epsilon_d$. The six remaining states
are each removed in energy by $U/2 \pm \epsilon_d$
or more, as detailed in Table~\ref{Table}. Defining
the operator
\begin{equation}
\hat{D} = c^{\dagger}_{1, 0} c^{\dagger}_{2, 0} d ,
\end{equation}
it is easy to verify that (i) $\hat{D}^{\dagger} \hat{D}$
projects onto the state $d^{\dagger} |0 \rangle$;
(ii) $\hat{D} \hat{D}^{\dagger}$ projects onto the state
$c^{\dagger}_{1, 0} c^{\dagger}_{2, 0} |0 \rangle$;
and (iii) $\hat{D}^2 = ( \hat{D}^{\dagger} )^2 = 0$.
In addition, $\hat{D}$ and $\hat{D}^{\dagger}$
anticommute with all electronic operators further
down the chain, namely, $c_{j, n}$ with $n \geq 1$.
Thus, $\hat{D}$ acts as a {\em conventional fermion
annihilation operator} within the truncated Hilbert
space where only the two lowest lying local states
are kept.

To eliminate the locally excited states, we perform a
canonical transformation akin to the Schrieffer-Wolff
transformation~\cite{SW66} for the Anderson model:
${\cal H}' = e^{S} {\cal H} e^{-S}$. Here the
anti-hermitian operator $S$ is chosen such that the
excited and low-lying sectors are decoupled within
${\cal H}'$ to linear orders in $t_1$, $t_2$, and
$\xi_0$. The explicit form of the operator $S$ is quite
cumbersome, and will not be specified here. Instead,
we proceed directly to the end result. Projecting
${\cal H}'$ onto the low-energy subspace and settling
with linear order in $1/U$, the effective low-energy
Hamiltonian reads
\begin{eqnarray}
{\cal H}' &=&
       \sum_{j = 1, 2}
       \sum_{n = 1}^{\infty}
             \xi_n \left \{
                            c^{\dagger}_{j, (n + 1)}
                            c_{j, n}^{} + {\rm H.c.}
                   \right \}
\nonumber \\
   &+& \epsilon_d \hat{D}^{\dagger} \hat{D} +
       \sum_{j = 1, 2}
             (-1)^{j} t'_j
             \left \{
                      c_{\bar{j}, 1}^{} \hat{D}
                      + {\rm H.c}
             \right \}
\nonumber \\
   &-& U' \left (
                 \hat{D}^{\dagger} \hat{D} - \frac{1}{2}
         \right )
         \sum_{j = 1, 2}
                 \left (
                         c^{\dagger}_{j, 1} c^{}_{j, 1}
                         - \frac{1}{2}
                 \right ) ,
\label{H'_latt}
\end{eqnarray}
where
\begin{equation}
t'_{j} = 4 \frac{t_{j} \xi_0}{U}
\label{t'_j}
\end{equation}
and
\begin{equation}
U' = 4 \frac{\xi_0^2}{U} .
\label{U'_latt}
\end{equation}
Here $\bar{j} = 3 - j$ marks the lead index opposite
to $j$. To linear order in $1/U$, the projected
nonequilibrium operator $Y'_0 = e^{S} Y_0 e^{-S}$
remains given by Eq.~(\ref{Y_0_latt}), apart
from the summation over $n$ which now runs over
$n \geq 1$. We emphasize that $Y'_0$ does acquire
corrections at order $1/U^2$, as does the energy
$\epsilon_d$ in Eq.~(\ref{H'_latt}). However, these
corrections are negligibly small when $U$ is large.

\begin{table}[tb]
    \centering
    \begin{tabular}{cc}
    \hline
    \hline
       \hspace{8mm} Eigenstate \hspace{8mm} &
             \hspace{8mm} Eigenenergy \hspace{8mm} \\
       \hline
       $|0 \rangle$ & $U/2$ \\
             \vspace{3pt}
       $d^{\dagger} |0 \rangle$
             & $-U/2 + \epsilon_d$ \\
             \vspace{3pt}
       $c^{\dagger}_{1, 0} |0 \rangle$ & $0$ \\
             \vspace{3pt}
       $c^{\dagger}_{2, 0} |0 \rangle$ & $0$ \\
             \vspace{3pt}
       $d^{\dagger} c^{\dagger}_{1, 0} |0 \rangle$
             & $\epsilon_d$ \\
             \vspace{3pt}
       $d^{\dagger} c^{\dagger}_{2, 0} |0 \rangle$
             & $\epsilon_d$ \\
             \vspace{3pt}
       $c^{\dagger}_{1, 0} c^{\dagger}_{2, 0} |0 \rangle$
             & $-U/2$ \\
             \vspace{3pt}
       $d^{\dagger} c^{\dagger}_{1, 0} c^{\dagger}_{2, 0}
            |0 \rangle$ & $U/2 + \epsilon_d$ \\
    \hline
    \hline
    \end{tabular}
    \caption{Eigenstates and eigenenergies of the local
             problem, defined by the Hamiltonian terms
             $\epsilon_d$ and $U$ in Eq.~(\ref{H_latt})
             (i.e., the couplings $t_1$, $t_2$, and
             $\xi_0$ are all set to zero). Here
             $|0 \rangle$ denotes the empty state with
             no particles available. For large $U$, the
             local spectrum consists of two low-lying
             state: $c^{\dagger}_{1, 0}
             c^{\dagger}_{2, 0} |0 \rangle$ with energy
             $-U/2$, and $d^{\dagger} |0 \rangle$ with
             energy $-U/2 + \epsilon_d$. The six
             remaining states involve an excitation
             energy of $U/2 \pm \epsilon_d$ or more.}
    \label{Table}
\end{table}

The structure of Eq.~(\ref{H'_latt}) is similar, but
not identical, to that of Eq.~(\ref{H_latt}). It
differs in the form of the $t'_j$ hopping terms, and
in the sign of the $U'$ interaction term. Both
differences are conveniently accounted for by the
particle-hole transformation
\begin{equation}
\tilde{c}^{\dagger}_{j, n} = (-1)^{j + n + 1}
                             c^{}_{\bar{j}, n} ,
\label{ph-trans}
\end{equation}
which mirrors the transformation
$\Phi_c(x) \to -\Phi_c(x)$ used in bosonization. In
this manner, $H'$ and $Y'_0$ regain the forms of
Eqs.~(\ref{H_latt}) and (\ref{Y_0_latt}) modulo
three differences. First, the summation over $n$ now
runs over $n \geq 1$. Second, $d^{\dagger}$ has been
replaced in Eq.~(\ref{H_latt}) with $\hat{D}^{\dagger}$,
which couples to $\tilde{c}_{j, 1}$ instead
of $c_{j, 0}$ ($j = 1, 2$). Third, the couplings
$t_{j}$ and $U$ have been replaced with $t'_{j}$
and $U'$ of Eqs.~(\ref{t'_j}) and (\ref{U'_latt}),
respectively. The same differences carry over to
the main observables of interest, namely, the
occupancy of the level and the current outgoing
from lead $j$. Explicitly, the corresponding
operators transform according to
\begin{equation}
\hat{n}'_d = e^{S} \hat{n}_{d} e^{-S}
           = \hat{D}^{\dagger} \hat{D} ,
\end{equation}
\begin{equation}
\hat{I}'_{j} = e^{S} \hat{I}_{j} e^{-S}
        = i(e/\hbar) t'_{j}
          \left [
                  \hat{D}^{\dagger} c_{j, 1}^{}
                  - c^{\dagger}_{j, 1} \hat{D}
          \right ] ,
\end{equation}
where we have again settled with linear order in
$1/U$ and with projection onto the low-energy
sector.~\cite{Comment-on-U^2} Thus, the nonequilibrium
IRLM defined by Eqs.~(\ref{H_latt}) and (\ref{Y_0_latt})
has been mapped for large $U$ onto a weakly interacting
version of the same model. This includes the
physical content of the operators $\hat{I}'_{j}$
and $\hat{n}'_d$ describing the current and
occupancy of the level.

Evidently, the nonequilibrium IRLM possess a
strong-to-weak-coupling duality also when placed on
a lattice. However, in contrast to bosonization,
mapping of large to small $U$ is only approximate,
being controlled by the small parameter $\xi_0/U \ll 1$,
and involves a significant reduction of the effective
tunneling matrix elements according to Eq.~(\ref{t'_j}).
On the other hand, the renormalized interaction
strength of Eq.~(\ref{U'_latt}) is in good
qualitative agreement with the predictions of
bosonization, which follows from the fact that
$\xi_0 \sim D \sim 1/\pi \rho$ for conventional
lattice models.

Our discussion thus far has focused on identical
leads with $\epsilon_n = 0$ along the chain. We
conclude this section by briefly commenting on
the case where either the $\epsilon_n$'s are not
identically zero in Eq.~(\ref{H_latt}) (marking
departure from particle-hole symmetry), or if the
leads have different underlying band structures
[i.e., $\epsilon_n \to \epsilon_{j, n}$ and $\xi_n
\to \xi_{j, n}$ in Eq.~(\ref{H_latt})]. In the most
general case, Eqs.~(\ref{t'_j}) and (\ref{U'_latt})
are replaced with
\begin{equation}
t'_{j} = 4 \frac{t_{j} \xi_{\bar{j}, 0}}{U}
\end{equation}
and
\begin{equation}
U' \to U'_{j} = 4 \frac{(\xi_{\bar{j}, 0})^2}{U} ,
\end{equation}
while $\epsilon_d$ acquires the additional shift
$\epsilon_d \to \epsilon_d - \epsilon_{1, 0} -
\epsilon_{2, 0}$. Here $\bar{j} = 3 - j$ is the
lead index opposite to $j$. Hence, the capacitive
coupling $U'$ becomes lead dependent when the leads
have different structures. Another crucial point
pertains to the particle-hole transformation of
Eq.~(\ref{ph-trans}), which maps each lead onto
a particle-hole reflected image of its partner.
Consequently, the structure of each lead is generally
altered by the mapping. These modifications, however,
do not change the main physical result, namely, that
the nonequilibrium IRLM with large $U$ is mapped onto
a weakly interacting variant of the same model.

\section{Physical properties for large $U$}
\label{sec:expressions}

Focusing on the limit where $U$ is large, we next
exploit the duality between strong and weak coupling
to derive explicit analytic expressions for the main
physical quantities of interest. These include the
low-energy scale $T_0$, the differential conductance,
$G$, and the occupancy of the level, $n_d$.

The scale $T_0$ is extracted from the equilibrium
Hamiltonian ${\cal H}$, with the voltage bias set to
zero. Starting from the dual Hamiltonian ${\cal H}'$,
the latter can easily be brought to the canonical form
considered in Ref.~\onlinecite{BVZ07} by converting
to the ``bonding'' and ``anti-bonding'' combinations
\begin{eqnarray}
\psi_{b}(x) &=&
        \frac{t_{1}}{t'} \psi_{1}(x)
        + \frac{t_{2}}{t'} \psi_{2}(x) ,
\label{bonding} \\
\psi_{a}(x) &=&
        \frac{t_{2}}{t'} \psi_{1}(x)
        - \frac{t_{1}}{t'} \psi_{2}(x) .
\label{anti-bonding}
\end{eqnarray}
Here
\begin{equation}
t' = \sqrt{ t_1^2 + t_2^2}
\label{t'}
\end{equation}
is the tunneling matrix element between the level
and the ``bonding'' band. No tunneling takes place
between the level and the ``anti-bonding'' electrons,
only capacitive coupling is left.
Equations~(\ref{bonding})--(\ref{t'}) pertain to
the dual Hamiltonian obtained within bosonization.
A similar construction applies to the lattice
version of ${\cal H}'$, in which case Eq.~(\ref{t'})
is replaced with
\begin{equation}
t' = \frac{4 \xi_0}{U} \sqrt{ t_1^2 + t_2^2}
\end{equation}
[see Eq.~(\ref{t'_j})]. Note that the conversion to
``bonding'' and ``anti-bonding'' modes is a standard
transformation in equilibrium. However, it looses
its usefulness when a finite bias is applied as the
two modes couple within $Y'_0$.

In the notations of Borda {\em et al},~\cite{BVZ07}
the converted Hamiltonian corresponds to $N = 2$,
$V_0 = t'$, and $U = U'$. Hence, one can read off
the low-energy scale from the renormalization-group
equations derived by these authors, which give
\begin{equation}
T_0(\rho U \gg 1) =
         D \left(
                   \frac{\Gamma'}{D}
           \right )^{1 - 2\rho U'}
\label{T_0}
\end{equation}
with
\begin{equation}
\frac{\Gamma'}{\Gamma} =
    \left \{
             \begin{array}{cc}
             1 & {\rm Bosonization} \\ \\
             (4 \xi_0/U)^2
             & {\rm Lattice\; cutoff}
             \end{array}
    \right. \, .
\label{G'/G}
\end{equation}
Here $U'$ is specified by Eqs.~(\ref{U'}) and
(\ref{U'_latt}) for the cutoff schemes used in
bosonization and on a lattice, respectively.
In writing Eqs.~(\ref{T_0}) and (\ref{G'/G})
for a lattice cutoff, we have implicitly assumed
that the effective bandwidth $D$ and density of
states $\rho$ are left unchanged upon converting
from $c_{j, 0}$ to $c_{j, 1}$.

In case of bosonization, Eq.~(\ref{T_0}) takes
the explicit form
\begin{equation}
T_0(\rho U \gg 1) =
         D \left(
                   \frac{\Gamma}{D}
           \right )^{1 - 8/(\pi^2 \rho U)} ,
\end{equation}
consistent with Eq.~(\ref{SBA-T0}) upon the simple
substitution $1/U_{\rm SBA} \to 4/U_{\rm bosonization}$.
This relation between the couplings defined within
the different cutoff schemes also showed up in the
form of the assumed duality in the SBA:
$\pi \rho U_{\rm SBA} \to 1/(\pi \rho U_{\rm SBA})$
as compared to $\pi \rho U \to 4/(\pi \rho U)$ in
bosonization. It is, however, easy to check that
this simple relation between the coupling constants
defined within the two schemes holds only
at strong coupling, and does not extend to weak
coupling. (The relation, in general, is nonanalytic,
and may include singular points.) Note that both
the SBA and bosonization expressions for $T_0$
approach $\Gamma$ as $U \to \infty$. Indeed, fixing
$\Gamma$ and increasing $U$, Eq.~(\ref{T_0}) reduces
asymptotically to $\Gamma'$, which equals $\Gamma$ in
case of bosonization. In contrast, $T_0(U)/T_0(0)$
equals $(4 \xi_0/U)^2$ for sufficiently large $U$
in case of a lattice cutoff. The latter result
agrees well with numerical renormalization-group
calculations,~\cite{Borda-private} where $\xi_0$
is given by the hopping matrix element between the
zeroth and first Wilson shells.

Since $U'$ is a marginal operator that drops to zero
as $U \to \infty$, the nonequilibrium IRLM reduces
asymptotically to a noninteracting level (albeit
with a strongly renormalized hybridization width
in case of a lattice cutoff). The effect of $U'$
is controlled by the dimensionless parameter
$\rho U' \ln (D/\Gamma')$. When made sufficiently
small, the differential conductance and occupancy
of the level assume standard noninteracting forms:
\begin{equation}
G = \frac{G_0}{2}
    \left [
            \frac{T_0^2}
            {(\epsilon_d + eV/2)^2 + T_0^2}
          + \frac{T_0^2}
            {(\epsilon_d - eV/2)^2 + T_0^2}
    \right ]
\end{equation}
and
\begin{eqnarray}
n_{d}(T, V) &=& \frac{1}{2} -
         \frac{1}{\pi} \frac{\Gamma_L}{\Gamma}
         \arctan
         \left (
                  \frac{\epsilon_d - eV/2}
                       {T_0}
         \right )
\nonumber \\
     &-& \frac{1}{\pi} \frac{\Gamma_R}{\Gamma}
         \arctan
         \left (
                  \frac{\epsilon_d + eV/2}
                       {T_0}
         \right ) ,
\end{eqnarray}
with
\begin{equation}
G_0 = \frac{e^2}{h} \frac{4 \Gamma_L \Gamma_R}
                         {(\Gamma_L + \Gamma_R)^2} .
\end{equation}
Here we have settled for brevity with zero temperature.
The extension to finite temperature is straightforward.
Note that the peak conductance, $G_0$, is independent
of the cutoff scheme, and is identical to its value
when $U = 0$. This follows from the fact that
$t'_R/t'_L$ is equal to $t_R/t_L$ in Eq.~(\ref{t'_j}).
This result for $G_0$ is in fact more general. It extends
to all values of $U$, as can be seen from a phase-shift
analysis.

\section{Conclusions}
\label{sec:conclusions}

The description of quantum impurities out of
equilibrium remains an outstanding
theoretical challenge, with an urgent need for
benchmark results. In this paper we provided such
a result, by showing that the nonequilibrium IRLM
with strong capacitive coupling is equivalent to
a weakly coupled version of the same model. When
expressed in terms of the proper low-energy scale,
the differential conductance and occupancy of the
level reduce for large $U$ to standard noninteracting
forms. This result is universal, independent of
the cutoff scheme used. Different cutoffs are
distinguished by the parametric form of $T_0$,
which is nonuniversal. In this respect there
appears to be a qualitative difference between
lattice cutoffs with a finite bandwidth, and
continuum-limit cutoffs of the type used in
bosonization and the SBA, where the high-energy
cutoff is sent to infinity. Within the latter
schemes, $T_0$ regains its noninteracting value,
$T_0(U = 0) = \Gamma$, as $U \to \infty$.
By contrast, $T_0$ decays to zero as
$T_0(U) \sim T_0(U = 0)/(\rho U)^2$ when placed
on a lattice. This difference must be kept in mind
whenever comparing different computational schemes.
On the other hand, the peak conductance is
independent of the cutoff scheme used, being
equal to its $U = 0$ value.

We further wish to emphasize that the
strong-to-weak-coupling duality reported in this
paper is a distinct property of the IRLM with two
screening channels, or leads. The IRLM with a single
screening channel has long been known to map onto
the anisotropic Kondo model.~\cite{Schlottmann80}
The limit $U \to \infty$ corresponds in this mapping
to $J_z \to \infty$ ($J_z$ being the longitudinal
Kondo coupling), while $\epsilon_d$ translates
to a local magnetic field. From the mapping one
clearly sees that the single-channel IRLM with
$\epsilon_d = 0$ flows to the same low-energy fixed
point whether $U$ is small or large. Namely, all
values of $U$ are physically equivalent at energies
much smaller than $T_0(U)$. However, unlike the
two-channel case, large and small $U$ are
inequivalent at energies comparable to and exceeding
$T_0(U)$. Indeed, repeating the strong-coupling
expansion of Sec.~\ref{sec:lattice-cutoff} for
the case of a single screening channel one
readily finds two low-lying local states:
$d^{\dagger} |0\rangle$ and $c^{\dagger}_0|0\rangle$.
As these states are occupied by a single electron
each, charge fluctuations are strongly suppressed
between the local complex and truncated chain at
energies far below $U$. As a result, formation of
a scattering center is driven entirely by local
dynamics when $U$ is large, in contrast to the
case where $U$ is small.

A far more dramatic effect is found when the number of
screening channels exceeds three. Here $U = 0$ and
$U \to \infty$ flow to different low-energy fixed
points, one ($U = 0$) corresponding to a strongly
hybridized level, and the other ($U \to \infty$)
to a free impurity.~\cite{Comment-on-impurity} The
strong-coupling and free-impurity phases of the
model are separated in this case by a finite-$U$
Kosterlitz-Thouless transition line, analogous to
the ferromagnetic-antiferromagnetic transition line
of the anisotropic Kondo model. A detailed analysis
of the IRLM with multiple screening channels will
be published elsewhere.~\cite{SBZA07}

We are grateful to L. Borda, E. Boulat, S.-P. Chao,
E. Lebanon, H. Saleur, A. Zawadowski, G. Zar\'and,
and in particular to Pankaj Mehta and Achim Rosch
for stimulating and enlightening discussions. A.S.
was supported
in part by the Center of Excellence Program of the
Israel Science Foundation. N.A. was supported in part
by the Binational Science Foundation, and by NSF grant
No. 421776. This work was initiated while N.A. was
a Lady Davis Fellow at the Hebrew University.


\begin{thebibliography}{9}
\bibitem{Comment-on-SBA-DOS}
         Note that $\hbar v_F$ was set to one in
         the SBA study of Ref.~\onlinecite{MA06},
         resulting in $\pi \rho U \to U/2$.
\bibitem{Comment-on-impurity}
         The impurity comprises in this case of the
         level $d^{\dagger}$, plus the two local lead
         degrees of freedom to which it is coupled
         (i.e., $c^{\dagger}_{j, 0}$ with $j = 1, 2$).
\end{thebibliography}
\end{document}